\documentclass[12pt,english]{article}
\usepackage[utf8]{inputenc}
\usepackage{bm}
\usepackage{setspace}
\usepackage[round]{natbib}
\usepackage{eqnarray,amsmath}
\usepackage[pdftex]{graphicx}
\usepackage{pdfpages}
\usepackage{lineno}
\usepackage{float}
\usepackage{geometry}
\usepackage{hyperref}
\usepackage{listings}
\usepackage{animate}

\title{A Dynamic Theory of the Area of Distribution}

\author{Jorge Sober\'on$^{1}$,  \and Luis Osorio-Olvera$^{1,2}$}

\date{%
    $^1$J. Sober\'on (\url{jsoberon@ku.edu}). Corresponding author. Biodiversity Institute and Department of Ecology and Evolutionary Biology,
    University of Kansas, 1345 Jayhawk Blvd., 
    Lawrence, KS 66045,USA, ORCID:0000-0003-2160-4148\\
    $^{1,2}$L. Osorio-Olvera (\url{luis.osorio@iecologia.unam.mx}).
    Laboratorio de Ecoinformática de la Biodiversidad, Departamento de Ecología de la Biodiversidad, Instituto de Ecología. Universidad Nacional Autónoma de México. CP: 04510, M\'exico, ORCID:0000-0003-0701-5398\\%
    \today
}

\begin{document}

\begin{spacing}{1.9}
\maketitle
\section{Acknowledgements} We are grateful to our many colleagues in the University of Kansas Niche Modeling Group for many lively and useful discussions on the topics of the paper. LOO acknowledges partial support by Consejo Nacional de Ciencia y Tecnología (CONACyT; postdoctoral fellowship number 740751; CVU: 368747) and PAPIIT IA203922 and Nancy Galvez (Instituto de Ecología, UNAM) for  technical support. LOO and JS acknowledge Blitzi Soberon for endless support.

\section{Abstract}
\subsection{Aims}
To propose and analyze a general, dynamic, process-oriented theory of the area of distribution.
\subsection{Location}
Mexico, Southern United States and the Caribbean
\subsection{Taxon}
Pierid butterflies
\subsection{Methods}
The area of distribution is modelled by combining (by multiplication) three matrices: one matrix represents movements, another niche tolerances, and a third, biotic interactions. Results are derived from general properties of this product and from simulation of a cellular automaton defined in terms of the matrix operations. Everything is implemented practically in an R package.
\subsection{Results}
 Results are obtained by simulation and by mathematical analysis. We show that the mid-domain effect is a direct consequence of dispersal; that to  include movements to Ecological Niche Modeling significantly affects results, but cannot be done without choosing an ancestral area of distribution. We discuss ways of estimating such ancestral areas. We show that, in our approach, movements and niche effects are mixed in ways almost impossible to disentangle, and show this is a consequence of the singularity of a matrix. We introduce a tool (the Connectivity-Suitability-Dispersal plot) to extend the results of simple niche modeling to understand the effects of dispersal.
\subsection{Main conclusions}
 The conceptually straightforward scheme we present for the area of distribution integrates, in a mathematically sound and computationally feasible way, several key ideas in biogeography: the geographic and environmental matrix, the Grinnellian niche, dispersal capacity and the ancestral area of origin of groups of species. We show that although full simulations are indispensable to obtain the dynamics of an area of distribution, interesting results can be derived simply by analyzing the matrices representing the dynamics.  

\begin{flushleft}
\section{Introduction}

Understanding the structure and dynamics of distributional areas is fundamental in biogeography and macroecology \citep{Brown1995, Gaston2009, Rapoport1975, Udvardy1969}. Most recent work on this subject has focused on modeling potential areas of distribution with correlational ``niche models" (ENM) \citep{Guisan2000a, Petersonetal2011} that are later projected in geography \citep{Guisan2000a} to obtain species distribution models (SDMs). 

  ENM is a practical methodology with many applications \citep{Guisan2013}. However, it estimates static expressions of the correlation between environmental variables and the occurrence of records of a species. These are regions with environments ``similar", in some sense, to those where the species has been observed. These regions can be interpreted as maps of the suitable environment for a species, but not generally as maps of its actual distribution. At least two other factors are required to indicate species presence: biotic favorability, and accessibility to dispersal from an original region \citep{Cain1944,Good1931, Petersonetal2011}. To illustrate this idea, a simple graphical tool, the BAM diagram, was introduced by Soberon and Peterson (\citeyear{Soberon2005}). This heuristic device suggests that the actual area of distribution of a species depends on the ``scenopoetic" (i.e., non-interacting) environmental variables, denoted by \textbf{A}, the area of accessibility given the dispersal capacities of the species, or \textbf{M}, and its biotic environment, or \textbf{B}. This is illustrated in Figure \ref{fig0}.

\begin{figure}[h!]
	\centering
	\includegraphics[scale=.5]{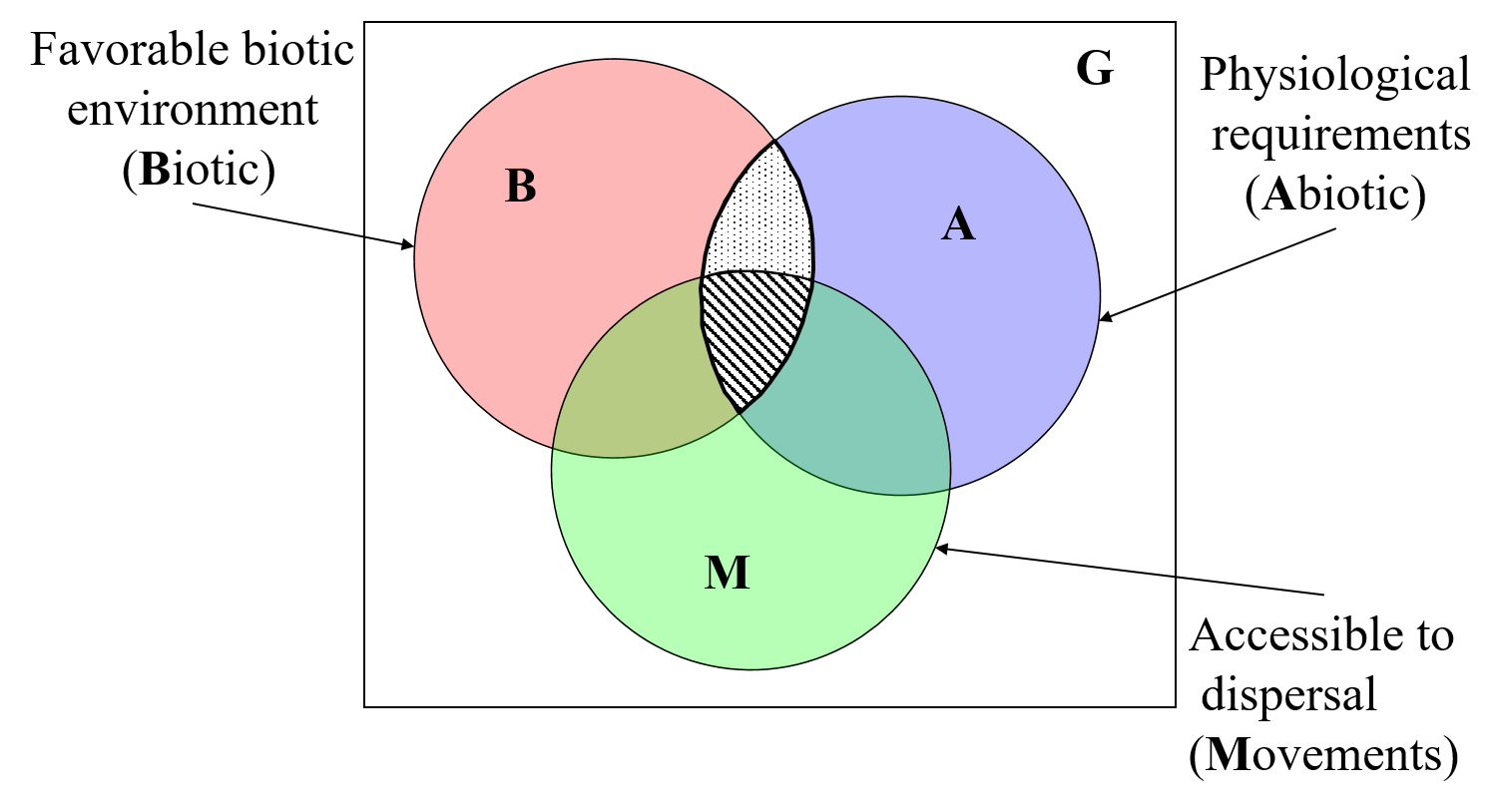}
	\caption{The BAM diagram. This illustrates the interaction of scenopoetic and biotic factors, with dispersal, that yield an actual area of distribution. The cross-hatched area is the actual area of distribution, and the dotted area a potentially habitable area.}
	\label{fig0}
\end{figure}

The simple BAM serve as a basis to establish dynamical equations for the area of distribution by combining niche models with movements and interactions in ``process-oriented models." Process-oriented models are dynamic and mechanistic, whereas conventional niche modeling is static and correlational. There are algorithms that include several of the mechanisms governing an area of distribution \citep{Cabral2017, Schurr2012a, Saupe2015, Rangeletal2007, Briscoe2019, Ovaskainenetal2002}, sometimes with a focus on parameter estimation \citep{Hooten2010b, Marion2012a, Hefleyetal2017MechanismMatter}. Moreover, many of the ideas of process-oriented models of the area of distribution have classic antecedents \citep{Skellam1951, Fisher1937}, or antecedents in metapopulation ecology \citep{HanskiGilpin1991, hanski1999metapopulation, Ovaskainenetal2002}.

In the future, process-oriented models are expected to play an important role in biogeography, and for this purpose we advance over the mainly heuristic BAM scheme \citep{Soberon2005,Soberon2017c} that provides an explicitly but static framework of the area of distribution, to provide a general, dynamic, process-oriented model of it. We use three classes of components in our scheme: first, realistic geographic and environmental scenarios; second, fundamental niches, dispersal capacities, and historical initial conditions for dispersal of each species (these are specie's traits); and, third, a system of equations relating the above. The modeling approach we used is a cellular automaton \citep{Wolfram2002}, a method that has been used in ecology to model spatially-explicit processes \citep{Molofsky2004}. Cellular automata not only emphasize simple rules leading to complex behavior, but also  require fewer parameters than demographic approaches \citep{Briscoe2019}. 

We focus on only two parts of the BAM: niche and dispersal, both of crucial importance to determine areas of distribution \citep{Ghergeletal2020}. Including biotic interactions is a substantial challenge due to complex theory and a lack of data, and it may be the case that interactions are less important at coarse geographical scales \citep{Petersonetal2011}. Our model is implemented with  practical R functions, capable of modeling distributions over tens of thousands of grid cells. Beyond the  practical tools provided, we also discuss theoretical insights derived from a mathematical analysis of the structure of the equation representing the BAM. This article represents developments over previous work \citep{Barve2011,Qiao2016}

\section{Theoretical basis}
Assume a geographical region of interest, with a grid of given extent, resolution, and map projection. Environmental values are associated with the region, as illustrated in figure \ref{fig:fig2}.

\begin{figure}[h!]

	\centering
	\includegraphics[scale=1]{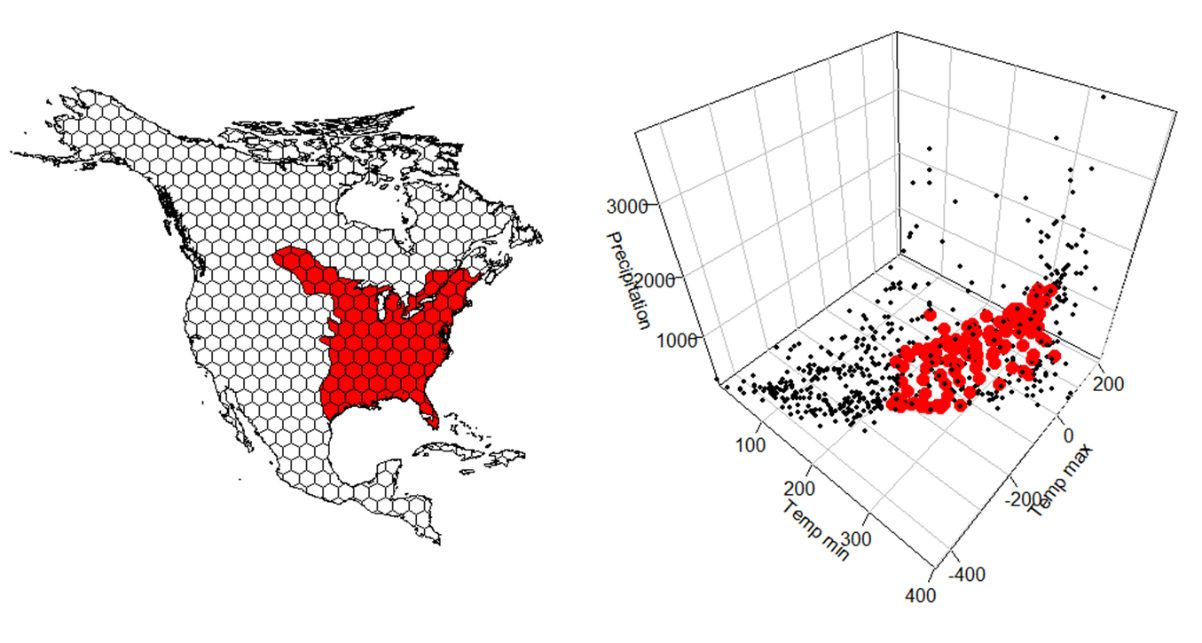}
	\caption{Extent of occurrence map for the squirrel \textit{Sciurus virginianus}, on a grid of 530 hexagonal cells of approximately 130 $km$ per side, and an example of environmental space, in three variables (minimum and maximum temperatures in centigrade X 10, and annual precipitation in mm). Albers equal area projection.}
	\label{fig:fig2}
\end{figure}

A simple representation of the area of distribution of a species (the red region in the figure) is a vector of binary values (530 in this case) expressing whether, at time $t$, cell $i$ is occupied by  species $j$. This is symbolized by the vector $\mathbf{G}_j(t)$. Although here we use a single species, we prefer to state the equations in general, for an entire group of species. Of course, a better description of the distribution might consist of continuous values representing population density, probability of presence, or a similar quantity, but such a description would require more detailed models, with many parameters that are difficult to measure.

In a binary scheme, a cellular automaton  would determine changes in the state of occupancy in every cell as a function of the connectivity of the cell (described below), and its environmental and biotic suitability \citep{Barve2011}. Starting with an initial configuration of zeroes and ones in the grid [$\mathbf{G}_j(0)$, at time $t=0$], the state in each cell may change or remain the same depending on its state at the previous time, and on whether there are new arrivals, the abiotic environment remains (or becomes) suitable, and the combination of species present is compatible with establishment. This can be modeled with a convenient combination of three matrix multiplications. 

If space is subdivided by a grid of $n$ cells (in the geographic plane, ie.e., cells with latitude and longitude coordinates), we first define an $n \times n$ matrix of adjacency \citep{Spielman2007} among all the cells in the grid, $\textbf{C}_j$, which we assume to be constant over time. Matrix $\textbf{C}_j$ expresses, for species $j$, which cells are \textit{adjacent} to which cells, regardless of environmental or biotic suitability, taking into account only dispersal. It is an expression of how connected cells are, given the dispersal capacity of the species. In other words, $\textbf{C}_j$ is species-dependent. Moreover, connectivity is also dependent on the length of the time step used, because, for instance, dispersing in a year is different from dispersing in a decade. We assume adjacency to first neighbors in one time-step, for low-dispersal capacity, and adjacency to second, third, or higher order neighbors for species with better dispersal capacity. 

The adjacency matrix is a very well-known object in graph theory, and we will base most of our results on this object, although other related and informative matrices (i.e., Laplacian, transition) exist \citep{Lovasz1993GraphsRandomWalks}. The Laplacian matrix approach would model dispersal as a fluid occupying the network of nodes, and the transition matrix approach as individuals exploring it in a random walk \citep{Briscoe2019,GilbertYeakel2019}. Our approach is similar to a ``contagion" process, whereby nodes can be visited from neighboring nodes, and, if conditions are right, become occupied. 

In general, and despite some limitations \citep{Moilanen2011}, the adjacency matrix scheme is a powerful method for representing movements, with deep theoretical meaning and results \citep{Spielman2007}. We are aware of the simplicity of basing the arguments merely on connectivity, rather than on actual dispersal, with all the complications of exponential decay, long-distance movements, and so forth \citep{Clobertetal2012}. However, despite their extreme simplicity, adjacency matrices are a useful tool to represent movements because they provide theoretical insights and are conceptually straightforward.

The second matrix that we use is derived from ENM calculations. We provide a threshold \citep{Liuetal2005} for the outputs of ENMs and create diagonal matrices with values of one corresponding to cells with model values above the threshold (meaning favorable environments), and of zero for model values below the threshold. This is denoted as $\mathbf{A}_j(t)$, an  $n \times n$ matrix representing the scenopoetic niche requirements of a species, at a given time. In general, $\mathbf{A}_j(t)$ is singular (something with implications for estimating ancestral areas, as discussed later), since it is diagonal with zero values in the unsuitable grid-cells . It may change much faster than the $\mathbf{C}_j$ matrix over time, since it is mainly determined by climate, which changes in time periods measured in decades \citep{Parmesan2003a}. $\mathbf{A}_j(t)$ is expected be obtained from fundamental niches, and thus ENM algorithms that  produce simple, convex shapes that reasonably resemble response surfaces may be preferred to complex methods better suited to fit  patterns affected by sampling-effort and other factors  \citep{Jimenezetal2019FundNich}. 

The third matrix would also be a diagonal matrix [$\mathbf{B}_j(t)$] with binary values. A value of one would indicate a biotic environment (of competitors, predators and mtualists) that allows a species to invade and persist in a given cell, and a value of zero the presence of interactors incompatible with invasion of species $j$ to cell $i$. Matrix $\mathbf{B}_j(t)$ will have very fast dynamics, commensurate with the time step of the simulation. Because we do not have enough data, we assume that this is the identity matrix. The general model then becomes: 
\begin{equation}
\mathbf{G}_j(t+1) =\mathbf{B}_j(t)\mathbf{A}_j(t) \mathbf{C}_j  \mathbf{G}_j(t)
\label{eq:automata}
\end{equation}
which, because we assume that biotic interactions can be disregarded (the matrix $\mathbf{B}$ is identity), reduces to:
\begin{equation}
\mathbf{G}_j(t+1) =\mathbf{A}_j(t) \mathbf{C}_j  \mathbf{G}_j(t)
\label{eq:automata2}
\end{equation}

The equation describes a very simple process: To find the occupied patches in $t+1$ start with those occupied at time $t$ denoted by $\mathbf{G}_j(t)$, allow the individuals to disperse among adjacent patches, as defined by $\mathbf{C}_j$, then remove individuals from patches that are unsuitable, as defined by $\mathbf{A}_j(t)$.  Equation \ref{eq:automata2} is formally that of a cellular automaton, which can be studied by simulation, as we do later, but several interesting properties of the model may be deduced from the matrix structure of the operations, as discussed in the appendix. Our approach is akin to classic metapopulation models \citep{Ovaskainenetal2002, hanski1999metapopulation, HanskiGilpin1991}, but we prefer the simpler cellular automaton approach because its logic is very transparent and because it is feasible to parameterize it at the large spatial and temporal extents typical of biogeography.

\section{Methods}

As a first region of study, we use Mexico, the Central American countries and part of the southern USA. We partitioned the region with a raster at a resolution of $0.16666 ^{\circ}$, or 10'. This means a total of 7,364 cells (and the adjacency matrices having $7364 \times 7364$ cells, many of which will not be used by the software). As a second region we used the Great Antilles, at a resolution of 5', meaning 2,563 cells of $0.08333 ^{\circ}$.

We estimate the matrix $\mathbf{A}$ with ENM on data obtained from the Global Biodiversity Information Facility (GBIF), for the Mexican butterfly (\textit{Dismorphia amphione lupita}). This species was selected for illustrative purposes only. Suspect records were removed and, in order to reduce spatial biases, data was thinned using the R package NicheToolBox \citep{Osorioetal2020ntbox, Aielloetal2015Thin}. 


The same package was used to fit 95\% minimum volume ellipsoids, similar to Mahalanobis-distance niche models \citep{Farber2003a}. To obtain binary suitability matrices, thresholding of the continuous output of the niche model was performed \citep{Liuetal2005}, assigning a value of $0$ to cells  with outputs of the ENM in the lower 5\% of values in the occupied cells. Ellipsoid niche volumes were used because they are expected to be closer to fundamental niches than some more irregular shapes \citep{Jimenezetal2019FundNich, Drake2015Bagging}.

A new R package \texttt{bamm} was developed to create and manipulate the adjacency matrices for realistic-size grids (adjacency matrices with many millions of cells). The \texttt{bamm} package also allows for obtaining the spectral and other properties of these matrices, and simulation of the evolution of an area of distribution by using equation \ref{eq:automata2}. The operations are performed with sparse matrices, thus leading to a practical use of computer memory. The package is written in R, and the installation instructions can be found at the GitHub repository (\url{https://github.com/luismurao/bamm}).

To create an adjacency matrix, the functions \textbf{model2sparse} and  \textbf{adj\_mat} in the \texttt{bamm} package are used. The user provides a raster file with the extent and resolution required, and a number $d$ specifying how many pixels are ``adjacent" to every other pixel. In other words, adjacency matrices are dependent on the dispersal capacity assumed for a species. Adjacency uses the idea of the ``neighborhood of Moore" \citep{Gray2003OnWolfram} of a given cell, but at one, two..., $d$ neighbors. 

In general, for realistic extents and resolutions, truly gigantic matrices must be used; these are nevertheless manageable by the sparse matrix techniques implemented in \texttt{bamm}. As mentioned before, the adjacency depend on the number of neighbors regarded as contiguous. For instance, for the resolution of 10', at first neighbors, one is assuming that in one time-step, cells at $0.16666^{\circ} \times 1$ of distance can be visited. At second neighbors, it is assumed that at one step, cells within $0.16666 ^{\circ}\times 2$ can be visited, and so on. Examples are shown below.

The \texttt{bamm} package also allows calculation of the adjacency lists, eigenvalues and eigenvectors of the matrices, that can be used to display the properties of the BAM scheme.

We analized three extreme situations: one where the $\mathbf{M}$ matrix is identity, meaning that only niche limits the distribution; one where the entire region of interest is scenopoetically suitable, meaning only dispersal limits the distribution, and a combination where both niche and dispersal limitations affect the distribution.

\section{Empirical examples and results}
The results are organized depending on whether only dispersal, only scenopoetic niche (we disregard biotic interactions), or both are considered.

\subsection{Distribution limited by dispersal constraints}
The simplest case for equation \ref{eq:automata2} is the situation in which only movements matter, because the niche matrix $\mathbf{A}_j(t)$ is universally favorable. This situation is called the ``world of Wallace" by \cite{Owens2013}. Here, suitability is assumed to be equally favorable, and interacting species are absent or do not affect the dynamics. In other words, the entire process is governed by the adjacency matrix $\textbf{C}_j$, and both  $\textbf{A}_j(t)$ and $\textbf{B}_j(t)$ are identity matrices.

The first eigenvector of an adjacency matrix reveals how connected a grid is \citep{Straffin1980}, in the sense that it indicates the number of ways in which cells in the grid  can be visited from other cells. We display this eigenvector, geographically organized, for continental and for island adjacency matrices. We can assume that $\textbf{C}$ is essentially constant, at least for periods of time measured in thousands of years. 
The first area of interest is a grid with a resolution of 10' of arc (this produces a $\mathbf{C}$ of 54,228,496 cells of approximately 18.5 km of side in the mainland), for a region of the world centered in Mexico. First, we plot the connectivity on the basis of the $\mathbf{C}_j$. This is shown in figure \ref{fig:figMDMC}, in which a central region is highly connected, and the periphery of the region is much less-well connected. 

\begin{figure}[h!]
	\centering
	\includegraphics[scale=.9]{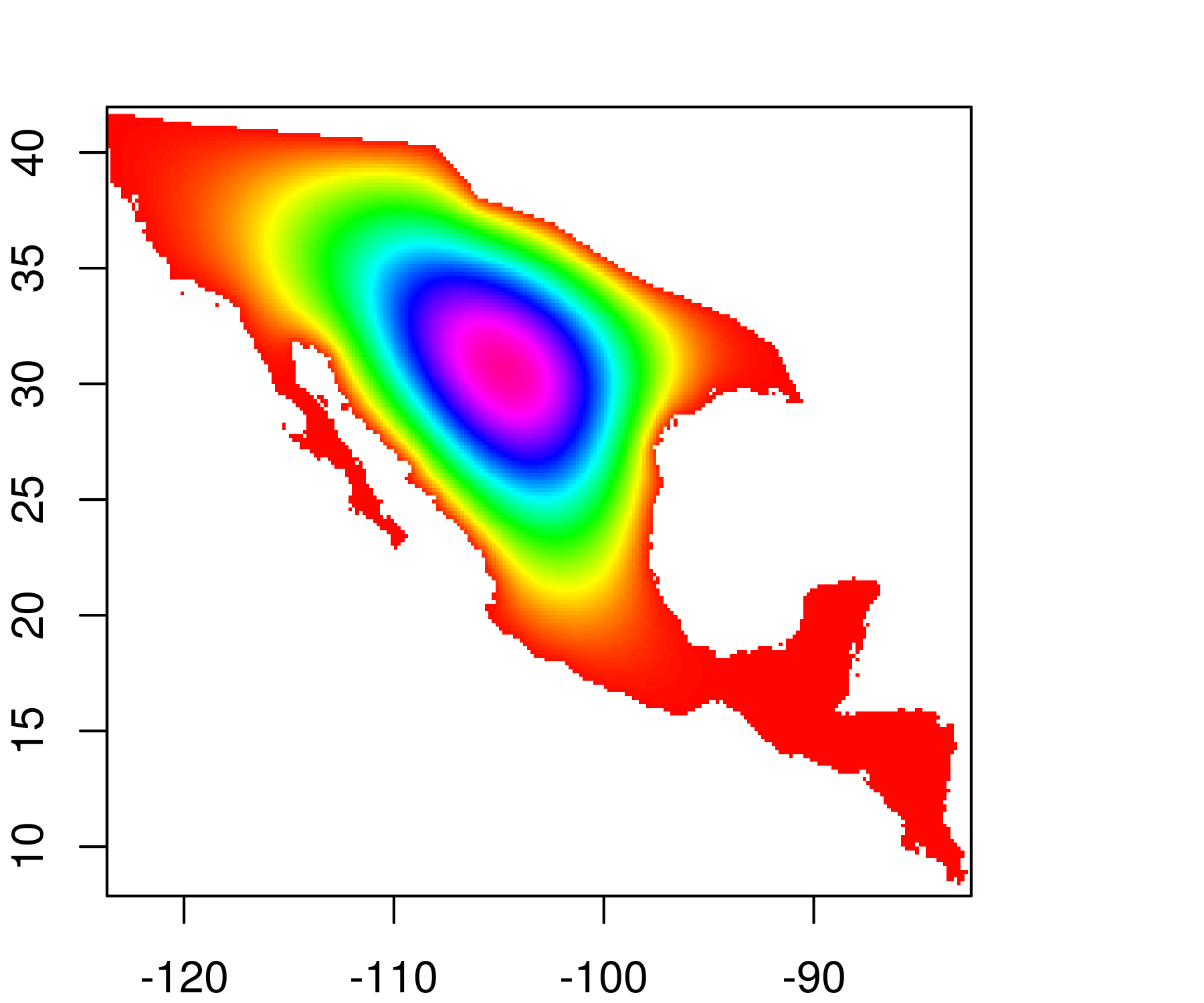}
	\caption{Frequency of visitation in a region.A mid-domain effect for a single species around Mexico. Red color indicates less frequent visits, and purple indicates high frequency of visits}
	\label{fig:figMDMC}
\end{figure}

Now we demonstrate the case of a region composed of islands. We use a resolution of 5' ($0.08333^{\circ}$, or approximately 9 km of side) but with different dispersal capacities. The results are shown in figs \ref{fig:AntillMay} and \ref{fig:ConAntillMay}

\begin{figure}[h!]
	\centering
	\includegraphics[scale=.35]{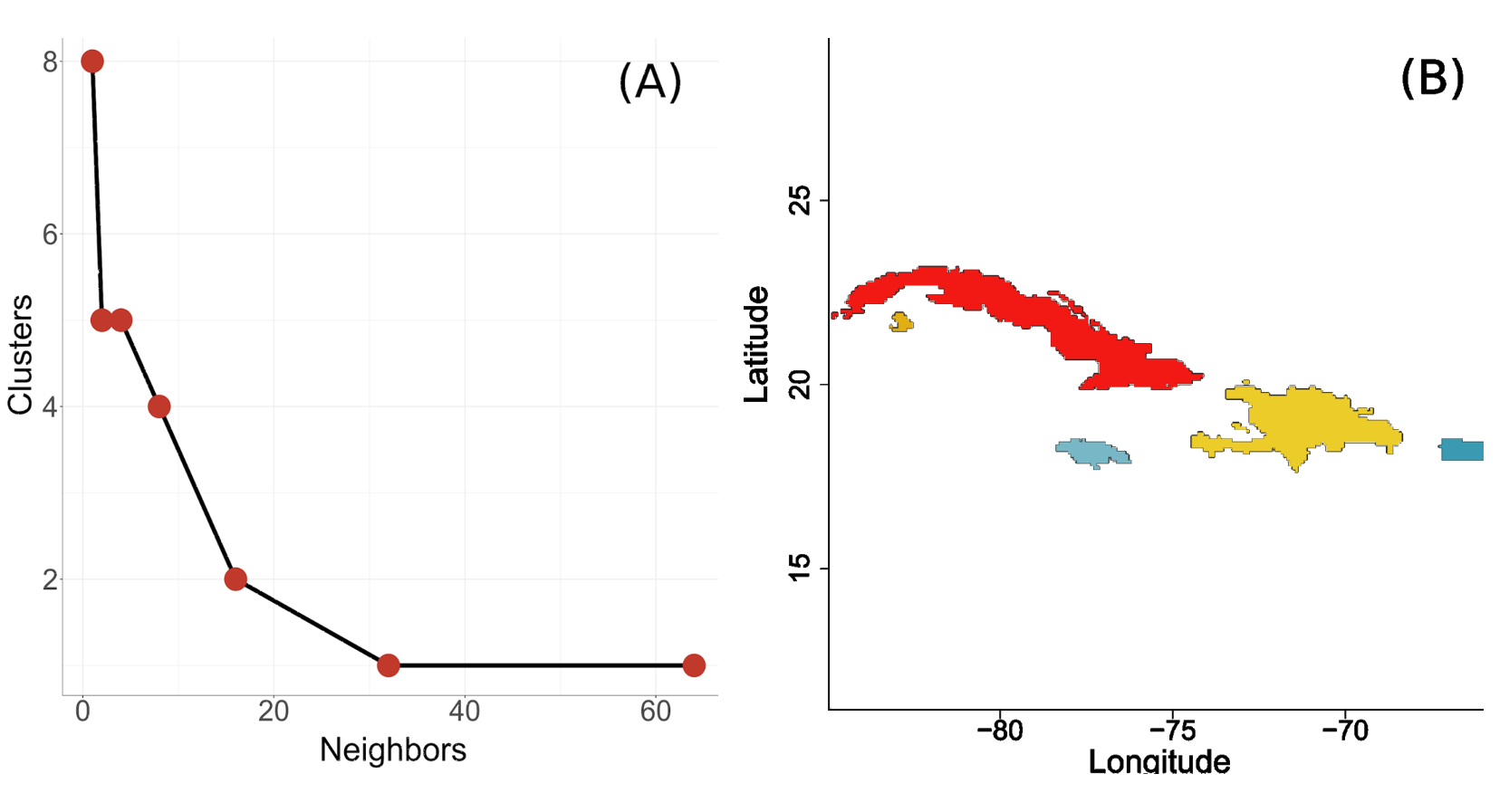}
	\caption{Structure of clustering of cells by dispersal parameter. (A) is a plot of the number of isolated clusters, as a function of how many pixels of $0.08333^\circ$ can be reached by migration, and (B) is the clustering at $d=4$ pixels of neighborhood.}
	\label{fig:AntillMay}
\end{figure}
\begin{figure}[h!]
	\centering
	\includegraphics[scale=.32]{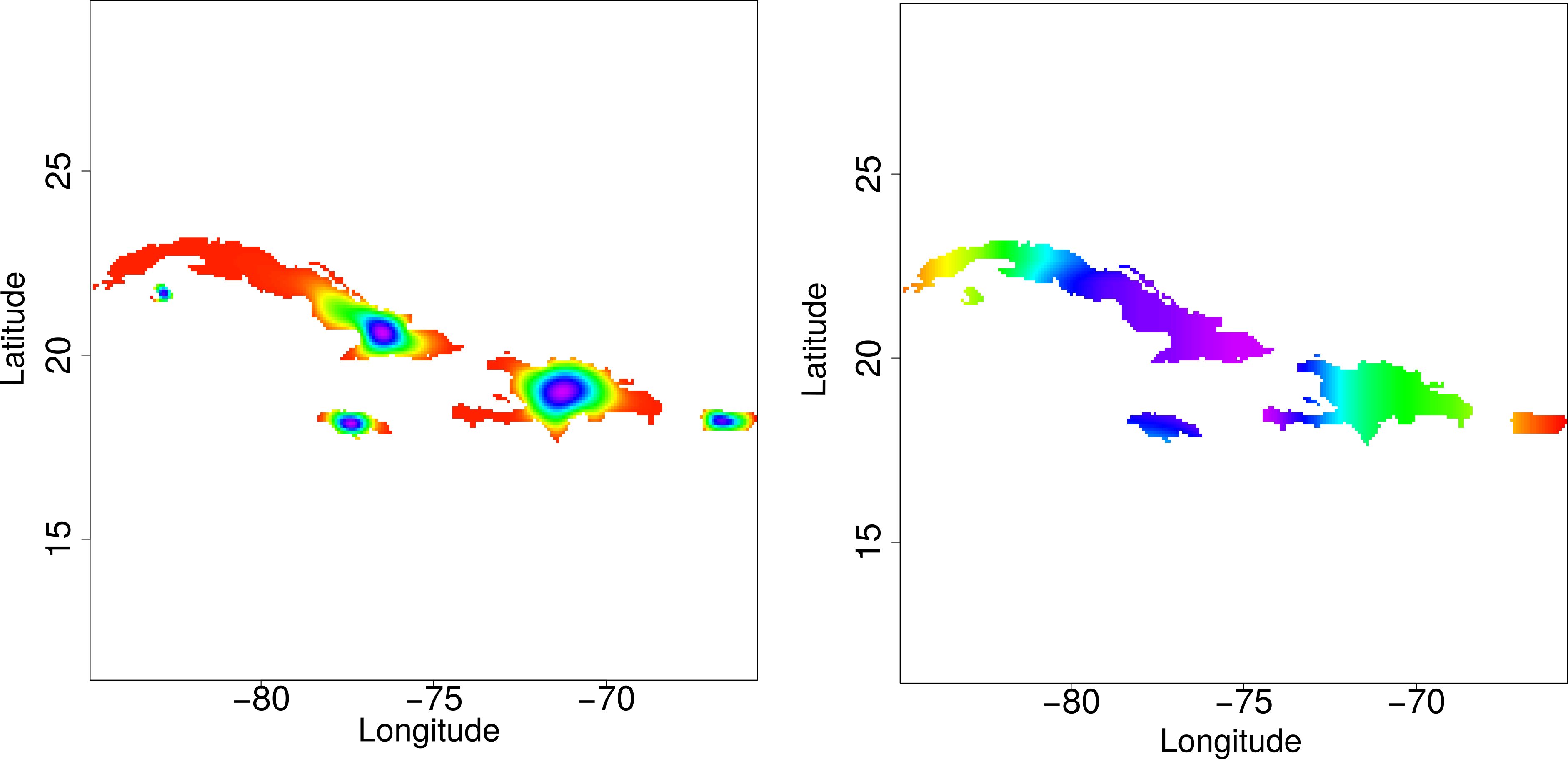}
	\caption{Connectivity of cells under contrasting values of $d$. (a) shows the connectivity structure for the $5$ clusters of pixels formed by assuming $d = 4 \times 0.08333 ^\circ$. (b) shows the connectivity structure for the single cluster at $d = 30 \times 0.08333^\circ$. Purple indicates a high level of connectivity, and red a low level.}
	\label{fig:ConAntillMay}
\end{figure}

The connectivity pattern in this fragmented landscape is different from the continental pattern in that there are fragments that constitute isolated clusters, if the value of $d$ is small enough. However, inside the clusters, the pattern of connectivity is similar to the continental pattern, with the center more often visited than the periphery. This is a (single species) pattern quite related to the mid-domain effect \citep{Colwell2000a} because by superimposing the patterns of many species, a full mid-domain effect can be obtained. As discussed later, this mid-domain effect is due entirely to the geometry of the region and to the dispersal properties of the species. 

In figure \ref{fig:ConAntillMay} we show a plot of how many clusters exist under different dispersal capacities ($d$). This informative plot will be generalized below to the case of connecting only suitable cells (as determined by an ENM). As presented, the information in this plot can be used to assess the dispersal capabilities required to connect the suitable parts of a fragmented landscape.
The above results correspond to the highly simplified case in which only dispersal capabilities matter. The results are interesting in the sense that they allow for replication of mid-domain effects on the basis of adjacency arguments, and also clarify that connectivity is depends both on geography and the species' dispersal capacities. In the Theory of Island Biogeography, dispersal, and thus colonization, become less likely at long distances, but the details of the actual geographical configuration (beyond distance and size) and differences in species environmental tolerances are disregarded. Our approach allows for these complications to be included, as shown below.

The world of Wallace includes only movements. We now consider a different case. 

\subsection{Distribution limited by scenopoetic niche constraints}
In a slightly more complicated case, only niche matters. A species will be present or not in a cell depending only on the scenopoetic suitability of the environment in the cell, disregarding movements. This extreme situation has been called Hutchinson's world \citep{Owens2013}. To calculate suitability, most researchers would use ENM, which estimates a potential area of distribution \citep{Franklin_Mapping_2010, Petersonetal2011,Guisan2017}. In this case, only the matrix $\textbf{A}_j(t)$ matters and the adjacency matrix $\textbf{C}_j$ connects everything to everything. As stated before, because the environments change, the matrix $\textbf{A}_j(t)$ also changes with climate, at speeds of perhaps decades \citep{Parmesan2003a}. 

The matrix $\textbf{A}_j(t)$ is defined on the basis of the composition of the environments in a given cell in relation to the those in cells where the species is present (assumed to be suitable), as shown in figure \ref{fig:DisNiche}.
\begin{figure}[h!]
	\centering
	\includegraphics[scale=.75]{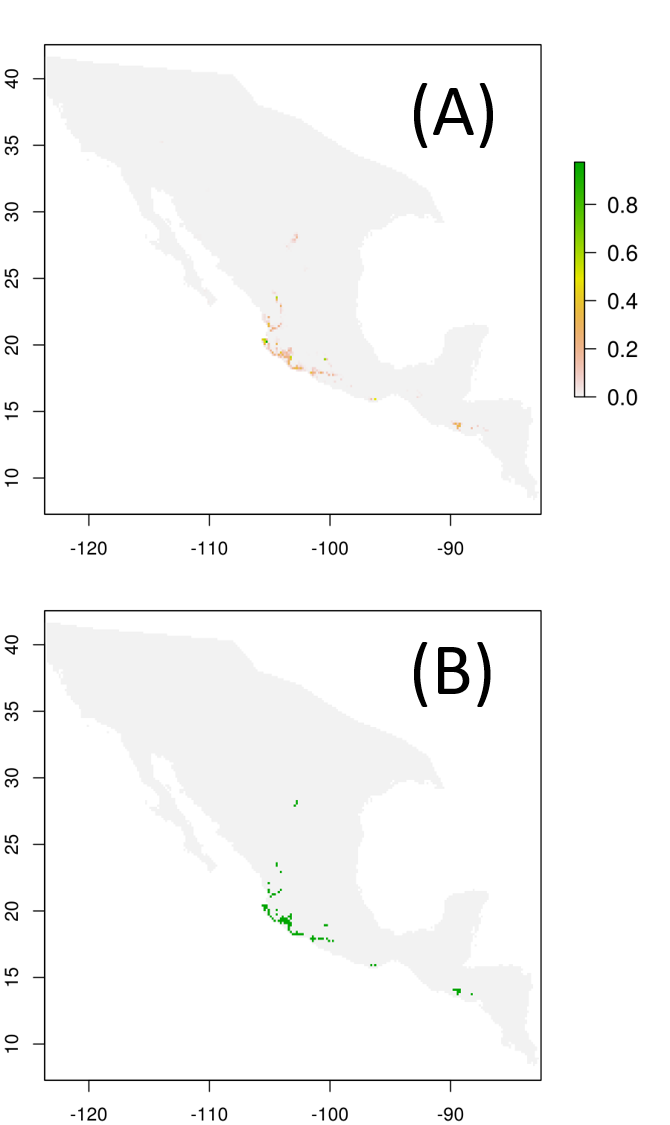}
	\caption{Suitability of cells for \textit{Dismorphia amphinoe lupita}. (a) shows a continuous measure of suitability, and (b) shows suitability thresholded to include cells from the bottom 5\% less suitable occurrences upwards (see methods).}
	\label{fig:DisNiche}
\end{figure}
A pure niche model as shown in the figure is simply a statement of the similarity of climates in a set of cells, to the climate in cells where a species has been reported to occur. Therefore, these models are simply ``\textit{niche}" models. To obtain an actual \textit{distribution} model (SDM), movements and biological interactions must be included, as described by equation \ref{eq:automata2}, and as performed next.

\subsection{Dispersal determined by niche and movements}
This is called "the world of Grinnell". In it, both movements and niche requirements (and eventually, biotic interactions) are combined. This is a much more complicated case, because now the dynamics is the result of the matrix product $\mathbf{A}_j(t)\mathbf{C}_j$, and this product combines movements and niche preferences in a manner nearly impossible to disentangle. The product is singular in general and thus it has no inverse. $\mathbf{A}_j(t)\mathbf{C}_j$ is a matrix representing connectivity from every grid-cell, to \textit{suitable} grid-cells. The eigenvector associated with its dominant eigenvalue shows how connected all the suitable cells are, as depicted in figure \ref{fig:AM}. However, ``connectivity" is species-dependent. An adjacency matrix of dispersal at longer distances would reveal connections amongst more grid-cells. Then, the plot of number of connected but isolated blocks of \textit{suitable} cells (including single cells), as a function of dispersal capabilities is very informative. The plot reveals the distance, assuming our simple model of adjacency, at which sets of suitable cells become connected. With far enough dispersal capacities, all cells form a single cluster. In the following example, we multiply the suitability matrix for \textit{Dismorphia amphione lupita} with connectivity matrices assuming dispersal capabilities to $1, 2, 10, 20, 30$, and $50$ neighbors of resolution of $0.16666^\circ$. The plot presented in figure \ref{fig:ScreePlotDa} shows how dispersal links isolated but suitable cells. We refer to this plot as a CSD (short of Connectivity Suitability Dispersal) plot. This graph complements a mere niche model, because it suggests the dispersal requirements to occupy isolated groups of suitable cells. However, as discussed below, to obtain the full information from equation \ref{eq:automata2}, initial conditions must be postulated, and a full simulation must be performed .

\begin{figure}[h!]
	\centering
	\includegraphics[scale=.35]{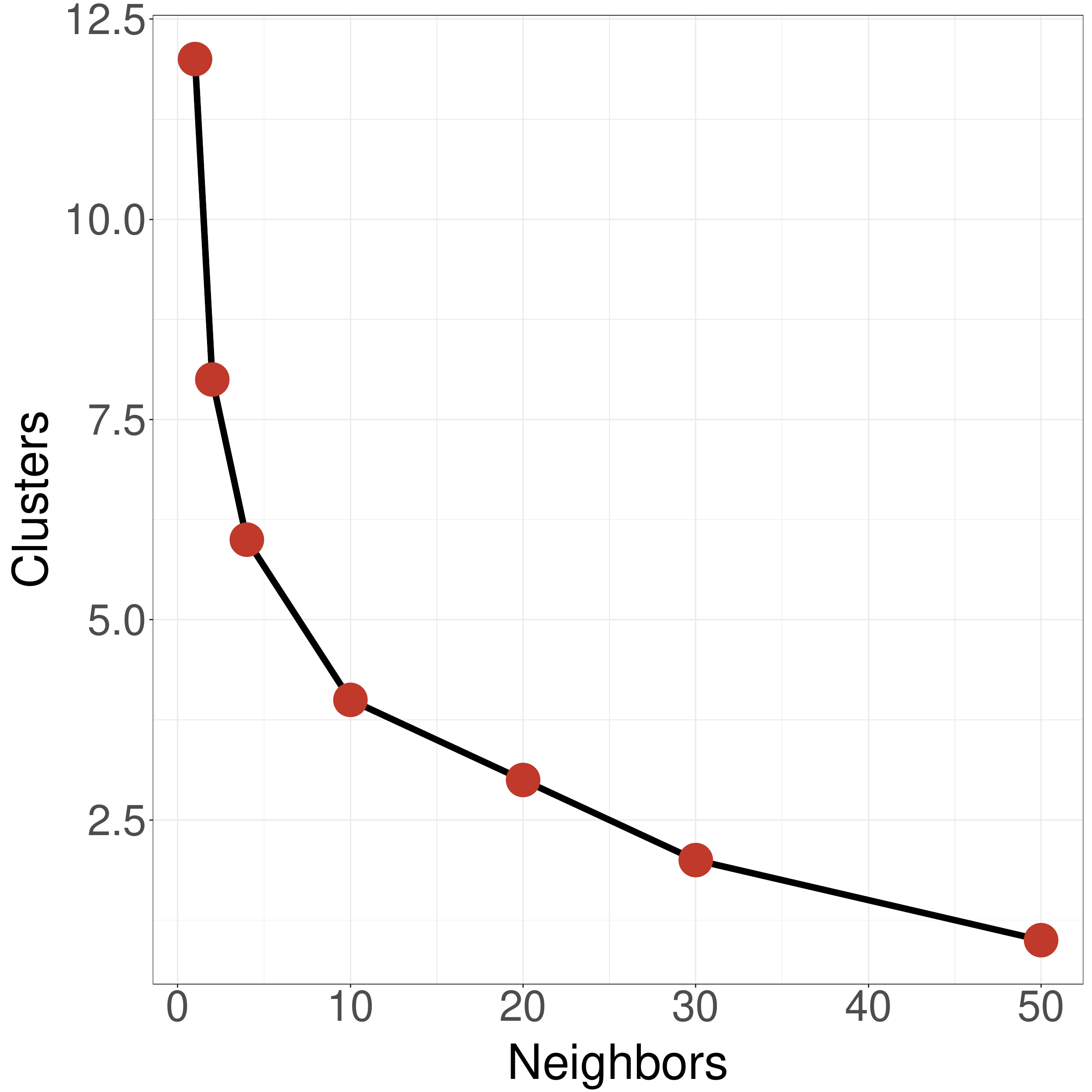}
	\caption{Number of connected clusters of suitable environment, as a function of dispersal (CSD plot), for \textit{Dismorphia amphione lupita} in Mexico}
	\label{fig:ScreePlotDa}
\end{figure}

\section{Climate change}
Under climate change, the niche matrix $\mathbf{A}_j(t)$ will also change, since it represents those cells with suitable climate under given fundamental niches, which are assumed to change very slowly \citep{PetersonNicheCons2011}. Using GCM data and one niche model, $\mathbf{A}_j(t)$ can be updated when climate changes. Therefore the spectral analysis of the product $\mathbf{AC}$ at different times would reveal important pattern changes. For instance, the average fragmentation over time can be studied as climate changes; this pursuit is interesting because more fragmentation might reasonably be assumed to be associated with higher speciation rates \citep{Qiao2017}. In figure \ref{fig:CSD_CC}, for the butterfly \textit{D. amphione lupita} we show CSD plots at three different times (last glacial maximum, mid Holocene, and present). 

\begin{figure}[h!]
	\centering
	\includegraphics[scale=.36]{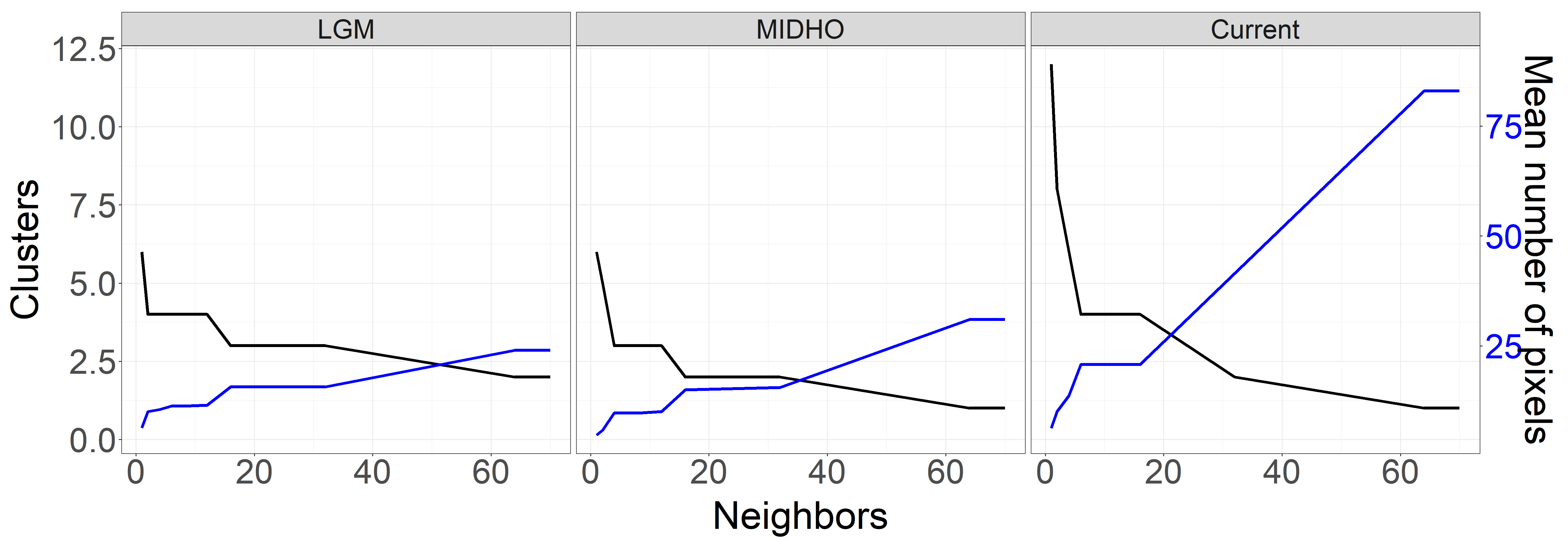}
	\caption{Connectivity for \textit{Dismorphia amphione lupita} in climate change scenarios: (A) last glacial maximum; (B) mid Holocene; (C) present.}
	\label{fig:CSD_CC}
\end{figure}

The plots shows that, in the past, the geographic expression of the niche of \textit{D. amphione} has been both smaller and less fragmented than today. Besides, because equation \ref{eq:automata2} can be used to simulate the development of an area of distribution, it is a straightforward procedure to model dynamically the effects of climate change on the distribution of a species, as shown in the Supplementary Materials.

\section{Simulations}

All the above results follow from applying graph-theoretical methods to link niche and dispersal. However, whenever the dispersal capacity of a species is small enough as to permit the existence of isolated regions of suitability, the origins of the dispersal must be hypothesized. By choosing different starting points, the results of the spread may be extremely different (see Supplementary Materials), and thus a full understanding of the distributional process requires a simulation approach \citep{Briscoe2019, Wolfram2002}.

Simulations highlight the difference between ``niche models", which are hypothesis about environmental tolerances of a species, and proper ``distribution models", which should be hypotheses about its geographical spread \citep{Soberon2017c}. In the Supplementary Materials, we present several animations displaying the operation of the model \ref{eq:automata2}. Although these simulations are indispensable as actual models of the dynamics of the distribution, they also have a theoretical interest, because they can be used to develop and add details to the original idea of a relationship between geographical and niche spaces \citep{Hutchinson1957a}. Moreover, in principle, $j=1, 2\dots, S$ species can be simulated, and then the metrics of biodiversity patterns, such as alpha and beta diversity, nestedness, and similarity, can be calculated \citep{Soberon2015b}. 

Simulations can also be applied, as described below, to the problem of assessing the likelihood of an ancestral area.

\section{Discussion}

This work combines dispersal with correlative ENMs to present a theoretical model of the area of distribution. There are some caveats to our approach. First, our model of dispersal as simple connectivity is limited. Assuming that cells are merely occupied or not, trough a process of contagion, does not take into account any of the complications of continuously diminishing probabilities of movement, or the crucial problem of fat-tailed distributions \citep{Clobertetal2012}. Moreover, this method of modeling dispersal ignores "viscosity" or "resistance" \citep{Gravesetal2014}. Barriers are established only by unfavourable niches.  Although this is an admittedly simplified model of dispersal \citep{BowlerBenton2005}, by using adjacency structures many results of graph theory are available, some of which we use. 

Second, many available methods allow for the estimation of the structure of a cluster of cells in geography \citep{Brownetal2017SDMToolbox, McGarigalMarks1995}, and have been used for a long time. There are active user communities familiarized with their implementation. These established methods can be used to produce a CSD plot. Our goal was not to introduce a new method for the analysis of spatial patterns, but rather to provide a dynamic, mathematical and computational version of the BAM diagram. Our emphasis is overtly theoretical, and some of our results can potentially be used to provide a conceptual framework for some indices that exist without substantial theoretical foundation. 

Third, a major simplifying assumption of our approach is that biotic interactions are disregarded. Of course, in general, interactions matter \citep{Schultz2022,Leathwick2007, Bullock2000a}, but including them requires not only data that are seldom available, but also to make decisions about very complex non-linear behavior \citep{Amarasekare2003}. A full treatment of biotic interactions in the scheme we presented is work for the future. 
\subsection{Some implications for biogeography}
The work presented here provides a conceptual structure that links, (1) realistic geographic reference zones (characterized by grids of given extent, resolution, and cartographic projection) and their corresponding environmental values; and (2), for a given pool of species of interest, their environmental requirements (also called ecological niches), their dispersal capacities, and their ancestral distributional areas. Although biotic interactions can be included in the scheme, as shown in equation  \ref{eq:automata}, in this work we ignore them, mostly because of a lack of data. 

Including the assumption of niche conservatism, the above scheme represents a null model for biogeography that relaxes the major assumption of unlimited dispersal capability used by Soberon (\citeyear{Soberon2019AmerNat}), and which was applied to the analysis of the species-area relationship. Including explicit dispersal capability and the historical ancestral origin of the species is a long-needed advance in modeling distributions of species. 

By considering dispersal explicitly, an interesting question can be addressed: what is the relative role of dispersal vs. niche breadth in driving  biodiversity patterns? The multiplication $\mathbf{A}_j(t) \mathbf{C}_j$ entangles the two factors. In explaining latitudinal gradients in species numbers, \citet{Saupe2019} concluded that both factors matter, and are modulated by the speed of climate change. In contrast, \citet{WormTittensor2018}  argue (on the basis of neutral and metabolic theories) that the signature of niche breadth is slight. Our matrix results clearly show that both factors should matter, because the characteristics of the niche may radically modify the dispersal possibilities and vice-versa, different dispersal capabilities produce different results for the same niche.

A classic topic in biogeography is the species-area relationship, or SAR \citep{ConnorMcCoy2013}. In some sense, increasing the value of $d$ in adjacency matrices is equivalent to increasing the area in a SAR, because with larger $d$ more area becomes available to a species. It would therefore be interesting then to analyze the SAR problem from the perspective of the position of fundamental niches in a growing niche-space, but adding limited dispersal capacities, in contrast to previous work \citep{Soberon2019AmerNat} in which unlimited dispersal is assumed.

As it is often the case when modeling by  cellular automaton \citep{Wolfram2002}, the rules for our scheme are simple and can be expressed as matrix multiplications. The simplest cases are modeled either by assuming differences only in connectivity among grid cells, which reduces to a mid-domain effect; or assuming niche differences and perfect connectivity, which reduces to niche modeling. In the most general case however, which we call  \textit{Grinnell's World}, both movements and niche requirements become entangled, such that distinguishing the factors is difficult, partly because the product matrix is singular, and its position in equation \ref{eq:automata2} makes it impossible to factor out. For the most cases then, understanding a particular distribution will require a full simulation, as has been discussed in general for cellular automata \citep{Wolfram2002}. However, we show that studying  some properties of the matrix $\mathbf{A}_j(t) \mathbf{C}_j$ is both practical and interesting, shedding light on the structure of connectivity of suitable cells, without requiring a full simulation to be performed.

A major result of our approach is the CSD plot. This plot reveals, for a given grid resolution, how distant are groups of connected and suitable grid cells. This result can be obtained in principle from the multiplicity of the eigenvalue zero of the Laplacian matrix associated with $\mathbf{A}_j(t) \mathbf{C}_j$ \citep{Lovasz1993GraphsRandomWalks}, but for large matrices, this represents a serious numerical problem. We therefore used the simple adjacency list of the suitable grid-cells, a computationally more tractable object. An empirical study of the structure of the CSD plots for a variety of species may yield valuable results and would eventually enable presence-absence matrices to be created for entire groups of species, on the basis of our scheme. This capability should provide the possibility for studying classic problems, such as Island Biogeography, from the novel perspective proposed in this paper.

The CSD plot has an antecedent in the work of Foltete et al. (\citeyear{Folteteetal2012GraphsSDMs}) who emphasize the need for adding connectivity arguments to geographically projected ENMs. Their work is different from ours in that they focus on landscape connectivity metrics as a tool to improve so called SDMs. In another antecedent, Peterson and Ammann (\citeyear{PetersonAmmann2013}) also suggested using fragmentation methods to analyze the outputs of niche models. The main difference between our work and theirs is that we introduce the CSD plot as a way to display the interaction between niche and dispersal features of a species, thus prompting the question of estimating ancestral areas of distribution.

Reconstructing ancestral distributions is a thriving field in systematics \citep{Ree2008} suggesting that our approach can be used for this purpose. However, our analysis reveals a problem. The matrix $\mathbf{A}_j(t) \mathbf{C}_j$ is in general singular. Because it has no inverse, the process described by equation \ref{eq:automata2} cannot be run backwards over time because when the projection matrix is singular, there may be different configurations that, after forward iteration, would lead to the same configuration, and thus, it is impossible to ``reverse" in a unique way the dispersal process. Therefore, in general, dispersal has a time arrow, and perhaps the only way of estimating ancestral areas of distribution is by assessing how likely a particular ancestral distribution is, given the present and the rules of the model. In other words, initial conditions are assumed, and some metric of map comparison is then used to obtain a distribution of differences between observed and simulated distributions. Initial conditions leading to the lowest differences are good hypothesis for an ancestral area.

Initial conditions could be proposed randomly, or perhaps using a CSD plot calibrated with past climate data (each cluster of suitable cells can be a postulated ancestral condition).  This method will work for a single species. For entire clades there are a number of methods available \citep{Clark2008} to estimate ancestral distributions. For a single species, our proposal shares the philosophy of searching, for groups of species, the most likely or most parsimonious ancestral area, by using a parametric model \citep{ReeSanmartin2009}. In our case, this would be equation \ref{eq:automata2}.  

The mid-domain effect is another biogeographical problem that can be clarified using \ref{eq:automata2}. The mid-domain effect is the idea that, if ranges of distribution would be placed randomly inside a region, there would be more overlaps in the central part of the region, and thus a larger number of species \citep{Colwell2000a} etc.). There has been a debate about whether expecting the effect is not only empirically true, but conceptually appropriate \citep{HawkinsDinizFilho2005}. Our work suggests that the mid-domain effect is simply the result of random movements with no environmental filtering and similar dispersal capabilities, rather than of randomly placing ``dispersal ranges" in an arena, a procedure of doubtful rigor \citep{HawkinsDinizFilho2005}. 

\subsection{Some implications for niche modeling}

So-called correlative distribution models are simply projections in space and time of a niche model. They are obtained by characterizing environments in a geographic grid on the basis of their similarity to environments in places where a species has been observed. They are practical and predictive \citep{Petersonetal2011} but disregard some crucial factors (movements and interactions) determining a distribution. We contend that to fully incorporate all factors, simulations must be performed. Many process-oriented models exist in this field \citep{Ghergeletal2020, Briscoe2019}, and one could ask why yet another algorithm should be introduced. Whether our approach is preferable, on empirical grounds, to any of the many methods \citep{Ghergeletal2020} that add movements to niche modeling  requires a comparison well beyond the mainly theoretical aims of this work. One of our major theoretical results is that the informative matrix product $\mathbf{A}_j(t) \mathbf{C}_j$ cannot be ``disentangled". Therefore, in general, a full simulation is required to establish a path of invasion. However, by using the \texttt{bamm} package, an adjacency matrix at a given $d$ can easily be obtained. Another one, on the basis of some geographic raster of a niche model, can then be multiplied to obtain the combination which yields information about what clusters of cell can be invaded, or even the full CSD plot for several $d$ values. This information is valuable for the analysis of invasive and migratory species, and for studies of climate change. We expect that combining ENMs with dispersal hypotheses will become common practice in our field \citep{Folteteetal2012GraphsSDMs}

In a related way, the CSD plot can be used to assess whether a given dispersal capacity is sufficient to link an entire projection of the niche model, in which case the final state of a simulation of the dispersal would not be sensitive to the initial conditions. However, if the hypothesis of the size of $d$ still predicts isolated clusters, then the dynamics of dispersal is sensitive to initial conditions. Limited dispersal capabilities require a hypothesis about the original source of dispersal. This is an extremely important point indicative of the deep differences between modeling niches (that can be projected to obtain \textit{potential} distributions), and modeling the dynamics of \textit{actual} distributions. For these, a combination of factors, including hypotheses about initial conditions must be incorporated. This point is almost universally ignored in the  distribution modeling literature. Nonetheless it is a well-known fact that outside a perfect Hutchinson's world, in which every part is accessible to every other part, clusters of suitable cells are generally disconnected, and therefore it is impossible for a dispersing organism to occupy a set of suitable cells in its entirety, unless the $d$ value is large enough, and dispersal equilibrium has been reached \citep{Svenning2004}. Estimating ancestral ranges thus becomes crucial for realistic simulations.

Another problem important in correlative niche modeling, is estimating the \textit{availability} region, or \textbf{M}, for a species. This should not be confused with the fact that some correlative niche modeling algorithms, such as MaxEnt \citep{Phillips2006b}, are sensitive to the choice of background. In ENM, \textbf{M} represents a region accessible to movements of a species. As such, the availability region \textbf{M} is an important concept in niche modeling \citep{Barve2011,Cooper2017}, regardless of the ENM algorithm used. \textbf{M} can be estimated by using the adjacency lists and the  CSD plot for the matrix $\mathbf{C}_j\mathbf{A}_j(t)$, which represents all the cells, suitable or not suitable, that can be reached from suitable cells, at a given dispersal parameter $d$. Establishing a cutoff value for $d$ provides a first order estimation of \textbf{M}. This can be improved by following the suggestion of Barve et al. (\citeyear{Barve2011}) and taking the union of suitability matrices at different times in the past. This technique is based on noticing that the matrices $\mathbf{A}_j(t)\mathbf{C}_j$ and $\mathbf{C}_j\mathbf{A}_j(t)$ have very different meanings, as explained in the appendix.

The fact that increasing $d$ means less fragmentation suggests an interesting prediction. The difference between the realized and the existing niches, as appears in the BAM diagram, would decrease as \textbf{M} grows (see figure \ref{fig0}). The prediction is that for species with high values of $d$, if there are differences between realized and projected (in geographic space) fundamental niches, these differences will depend mostly on biotic factors.

Finally, ENMs are also used in landscape genetics \citep{Maneletal2003} and phylogeography \citep{LunaArangureetal2020}. In both areas, an ENM complemented with information about dispersal distances and fragmentation of suitable patches would be valuable, because intrinsically those themes require information about how subpopulations are connected  \citep{Templetonatal1990Fragmentation}. The \texttt{bamm} package allows a CSD plot to be created directly from  the raster output of an ENM and hypothesis about dispersal, and we hope that increasingly, simple correlative ENM will be complemented by methods that allow dispersal to be included in the analyses.

\subsection{Conclusions}
In this paper, we presented a simple equation describing how the interplay between  niche and dispersal drives the dynamics of the area of distribution. This equation is based on classic ideas, and provides a fully-fledged computational theory of the area of distribution. We claim that although full simulations are required to take all the factors into account (specifically the ancestral region of origin of a species), a great deal of information and insights can be derived from the matrix operations representing the different factors. Our results are also practical, and we introduce an R package capable of handling the dynamics of the area of distribution, for large regions and at detailed resolutions. Finally, the work can be advanced along several lines, including better models of dispersal, and application to estimate regions of accessibility and ancestral areas. The fact that geographically realistic simulations can be performed is also useful to analyze the dynamics of invasive species.

\section{Appendix}
\subsection{Proof of the equivalence of two matrices of iteration}

Proof that $\mathbf{G}_j(t+1) = \mathbf{A}_j(t) \mathbf{C}_j(t)\mathbf{G}_j(t)$ is equivalent to $\mathbf{G}_j(t+1) = \mathbf{A}_j(t) \mathbf{C}_j(t) \mathbf{A}_j(t)\mathbf{G}_j(t)$. The result follows trivially from noting that the operation $\mathbf{A}_j(t)\mathbf{G}_j(t)$ simply places a zero value in every cell of $\mathbf{G}_j(t)$ with environments outside its fundamental niche. The remaining multiplications simply calculate movements to other cells and then allocates zeroes to every non-suitable cell. Therefore migrating from $\mathbf{G}_j(t)$ or from $\mathbf{A}_j(t)\mathbf{G}_j(t)$ is equivalent.

\subsection{Some useful properties}
\begin{itemize}
    \item $\mathbf{A}_j(t)=\mathbf{A}^T_j(t)$, which follows from $\mathbf{A}_j(t)$ being diagonal.
    \item $\mathbf{C}_j=\mathbf{C}^T_j$, which follows from $\mathbf{C}_j$ being symmetric
    \item
    $[\mathbf{A}_j(t) \mathbf{C}_j]^T =  \mathbf{C}_j(t) \mathbf{A}_j(t)$, which follows from the two preceding properties
\end{itemize}

\end{flushleft}
\subsection{Details of the \texttt{bamm} package}
The \texttt{bamm} package is an R package designed to create and operate on large (tens of millions of cells) matrices related to the BAM scheme, for instance, the adjacency matrix (connectivity matrix), and the niche suitability matrices. The package uses sparse matrices to represent those objects, thus allowing for efficient use of memory and small computation times.

The main function of the package are:
\begin{itemize}
    \item {\it \textbf{model2sparse}}: this is the basic function of the package. It converts a binary niche model (in raster format) to a sparse matrix model (object of class \textbf{setA}).
    
    
    \item {\it \textbf{adj\_mat}}: this function returns the sparse representation of the adjacency matrix of a given raster (generally is the \textbf{M} area but can be any area) given a movement hypothesis. The user can ask the function to return the eigen-analysis of the matrix. 
    
     \item {\it \textbf{bamm\_clusters}}: this function estimates the connectivity of suitable areas given an adjacency matrix. It returns three objects: a) an dynamic map (open-street map) of connected areas or clusters, b) the dataframe with coordinates of the geographic cluster membership, and c) a raster object of with cluster IDs.
     
      \item {\it \textbf{csd\_estimate}}: this function is used to estimate the CSD-plot shown in Figure \ref{fig:CSD_CC}.
      
       \item {\it \textbf{occs2sparse}}: this function converts occurrence data into a sparse matrix object. The object is used to declare the initial conditions for modeling the invasion dynamics of a species. 
       
       \item {\it \textbf{sdm\_sim}}: this function simulates single species dispersal dynamics by using the cellular automaton of the area of distribution (equation \ref{eq:automata2}). The parameters of the function are the sparse niche model,  an adjacency matrix, the initial points of invasion, and the number of simulation steps. The function returns a list of sparse matrices with the states of the cells (occupied or unoccupied) in each time step of the simulation.

\end{itemize}

In supplemental S1, we show functionalities to animate equation \ref{eq:automata2} and to display the results of the simulations as animated GIF files. 

\subsection{General matrix ideas}

The adjacency matrix $\mathbf{C}_j$ of the cells in a grid \citep{Spielman2007} represents an idealized world in which cells can be accessed from its neighbors. It represents an \textit{undirected} graph. We note that it is symmetric, because we assume that there are no preferred directions of movement, and if neighborhood is restricted to nearby cells, it is very sparse. Adjacency matrices of realistic grids are very large. If the grid has $n$ cells, the adjacency matrix is of size $(n \times n)^{2}$, and these large sizes have been used as an argument against using graph methods in ecology \citep{Moilanen2011}. However, adjacency matrices are practical as long as the appropriate numerical methods for sparse matrices are used (in this work we use R code for the purpose). 

In ecology, the spectral decomposition of the matrix of a graph is very informative \citep{DinizFilhoBini2005, LegendreFortin1989}. For instance, the eigenvector associated with the dominant eigenvalue of an adjacency matrix [which is strictly positive, \cite{Spielman2012,Straffin1980}] provides information about the number of forms in which a cell can be visited from other cells. As we saw, this is  directly related to the idea of the mid-domain effect \citep{Colwell2000a} for a single species. Indeed, overlaying the eigenvectors of the $\mathbf{C}_j$ matrices of a set of species will produce the mid-domain effect of a region, for a given set of species. In our approach, the adjacency matrix is given by the geometry of the region, and thus the dominant eigenvector for each species would be the same and the mid-domain effect follows directly. However, if one includes niche effects, then the eigenvectors of the product $\mathbf{A}_j \mathbf{C}_j$ are needed, and their overlaps can be extremely complicated. See the discussion.  
 
 In this work we will make use of the properties of spectral decomposition to advance our understanding of equation \ref{eq:automata2}, specifically of the connectivity matrix, and the product of connectivity and suitability.
 
 First, the sum of the squares of the eigenvalues of the matrix $\mathbf{C}_j$ (a symmetric matrix) is the number of edges in the graph. The $c_{i,h}$ entry in the $k-th$ power of $\mathbf{C}_j$ represents the number of ways in which node $h$ can be reached from node $i$ in $k$ steps \citep{Pavlopoulosetal2011}. All these results apply to the unrealistic case of total connectivity among cells in the grid, but they are interesting because they provide a ``null model" framework to understand more complex situations.

However, in equation \ref{eq:automata2}, we do not use $\mathbf{C}_j$, but $\mathbf{A}_j(t) \mathbf{C}_j$ (ignoring biotic interactions). This product induces an asymmetry, because now we have suitable and unsuitable nodes. In this case, the matrix $\mathbf{A}_j(t) \mathbf{C}_j$ represents a \textit{directed} graph, in which connections from unsuitable to suitable, but not vice versa, are allowed. Because it is the matrix of a directed graph, some of the spectral properties change, but we can use a trick to address this. First, $\mathbf{A}_j(t)\mathbf{C}_j$ is the asymmetric matrix of the directed graph of ``all connected to suitable" (as specified by $\mathbf{C}_j$), but it is easy to construct a symmetric matrix with similar properties. Indeed, $\mathbf{A}_j(t)\mathbf{C}_j\mathbf{A}(t)_j$ is the matrix of the undirected graph of ``suitable connected to suitable."  

\begin{figure}[h!]
	\centering
	\includegraphics[scale=.3]{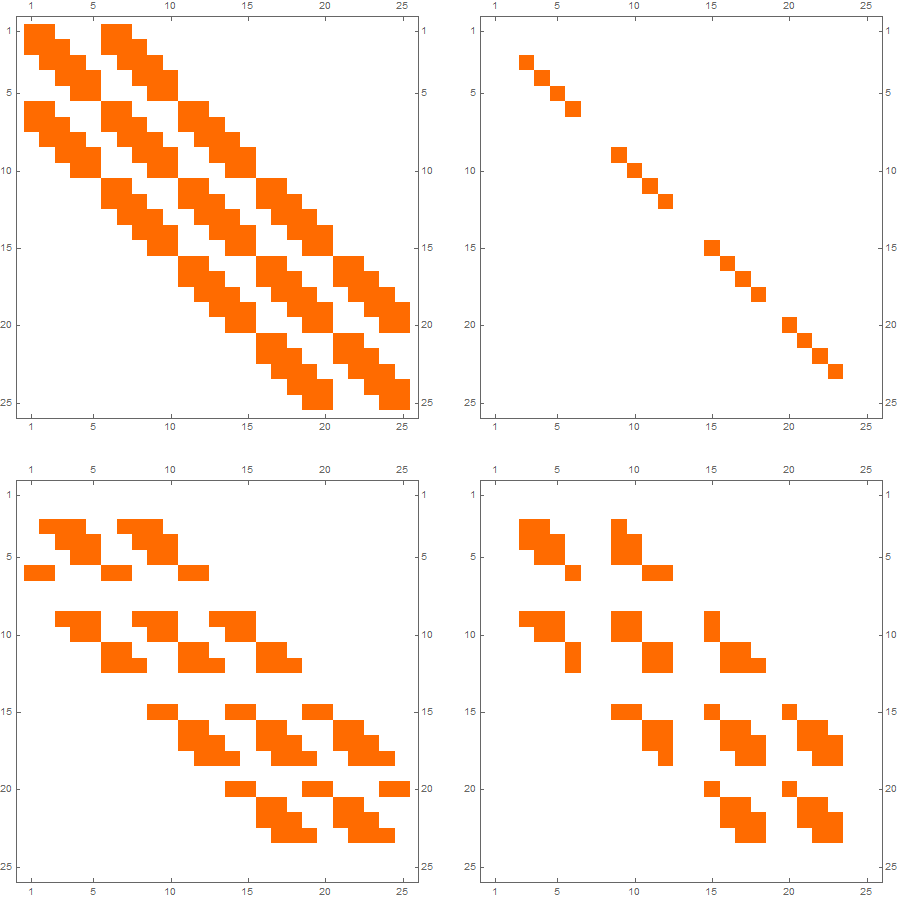}
	\caption{Connectivity of different $5 \times 5$ matrices. Top left, $\mathbf{C}_j$. Top right, a niche matrix $(\mathbf{A}_j$ with nine unsuitable cells. Bottom left, the product $\mathbf{A}_j(t) \mathbf{C}_j$. Bottom right, the product $\mathbf{A}_j(t) \mathbf{C}_j\mathbf{A}_j(t)$.Connected cells are shown in orange.}
	\label{fig:figMats}
\end{figure}

Before it was shown that the cellular automaton in equation \ref{eq:automata2}, using as iteration matrix $\mathbf{A}_j(t) \mathbf{C}_j$, is equivalent to one using $\mathbf{A}_j(t) \mathbf{C}_j\mathbf{A}_j(t)$. This result allows us to treat the matrix $\mathbf{A}_j(t) \mathbf{C}_j\mathbf{A}_j(t)$ as the representation of an undirected graph, with movements only among suitable patches. In figure \ref{fig:figMats}, we illustrate these ideas.

By multiplying the connectivity matrix by the niche matrix, a radical modification of the structure of connections is produced, as shown in an example in which the $5$ by $5$ grid, perfectly connected, is modified by a niche matrix with nine unsuitable cells. This is illustrated in figure \ref{fig:figMatsG} which shows the graphs of the different matrices. The perfectly connected grid at the top left is transformed, after multiplication of a niche matrix, into a set of two connected and suitable nets of cells, and nine non-connected cells  :
\begin{figure}[h!]
	\centering
	\includegraphics[scale=.3]{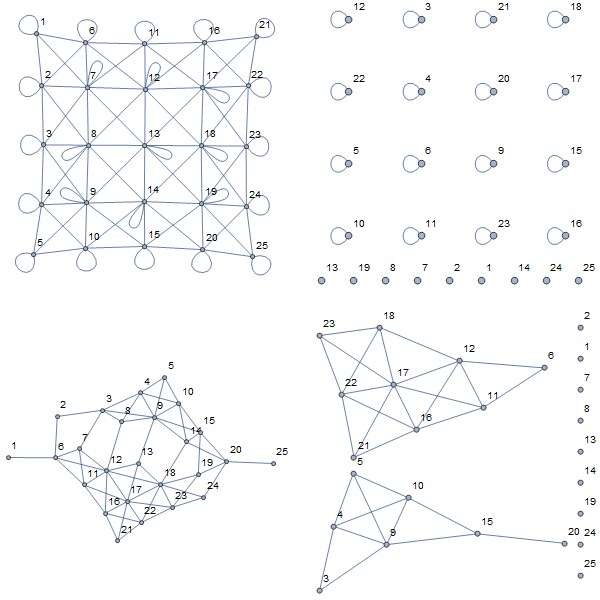}
	\caption{Four different graphs. Top left, completely connected grid of 25 cells. Top right, a grid with  with nine unsuitable cells. Bottom left, a graph of the product $\mathbf{A}_j(t) \mathbf{C}_j$. Bottom right, a graph of the product $\mathbf{A}_j(t) \mathbf{C}_j\mathbf{A}_j(t)$ showing two connected blocks of cells, and the nine unsuitable cells}
	\label{fig:figMatsG}
\end{figure}

We can then use the extremely useful result in which the multiplicity of the zero eigenvalue of the Laplacian matrix associated with $\mathbf{A}_j(t) \mathbf{C}_j\mathbf{A}_j(t)$ represents the number of \textit{connected components} of the grid. This is illustrated with a real-life example in the results section. 

The properties of the \textit{adjacency list},  \citep{Pavlopoulosetal2011} of the matrix $\mathbf{A}_j(t) \mathbf{C}_j\mathbf{A}_j(t)$ allow for calculation of the identity of cells connected to other cells, given a niche structure and  dispersal capacities. This was illustrated before using real species data.

Equation \ref{eq:automata2} (with no biotic interactions) suggests that the products  $\mathbf{A}_j(t) \mathbf{C}_j$ and $\mathbf{C}_j\mathbf{A}_j(t)$ are both interesting. Indeed, as mentioned before, the first product represents the cells in the grid that are suitable and can be reached in one step, given the connectivity structure. The second product represents those cells than can be reached from suitable cells, again given the connectivity structure. Both products have identical eigenvalues \citep{Williamson1954} but not necessarily equal eigenvectors. The first eigenvector of an adjacency matrix represents how connected every cell is to all others \citep{Pavlopoulosetal2011}, in the sense of how many ways exist for visiting each cell after a large number of dispersal steps.

\begin{figure}[h!]
	\centering
	\includegraphics[scale=.6]{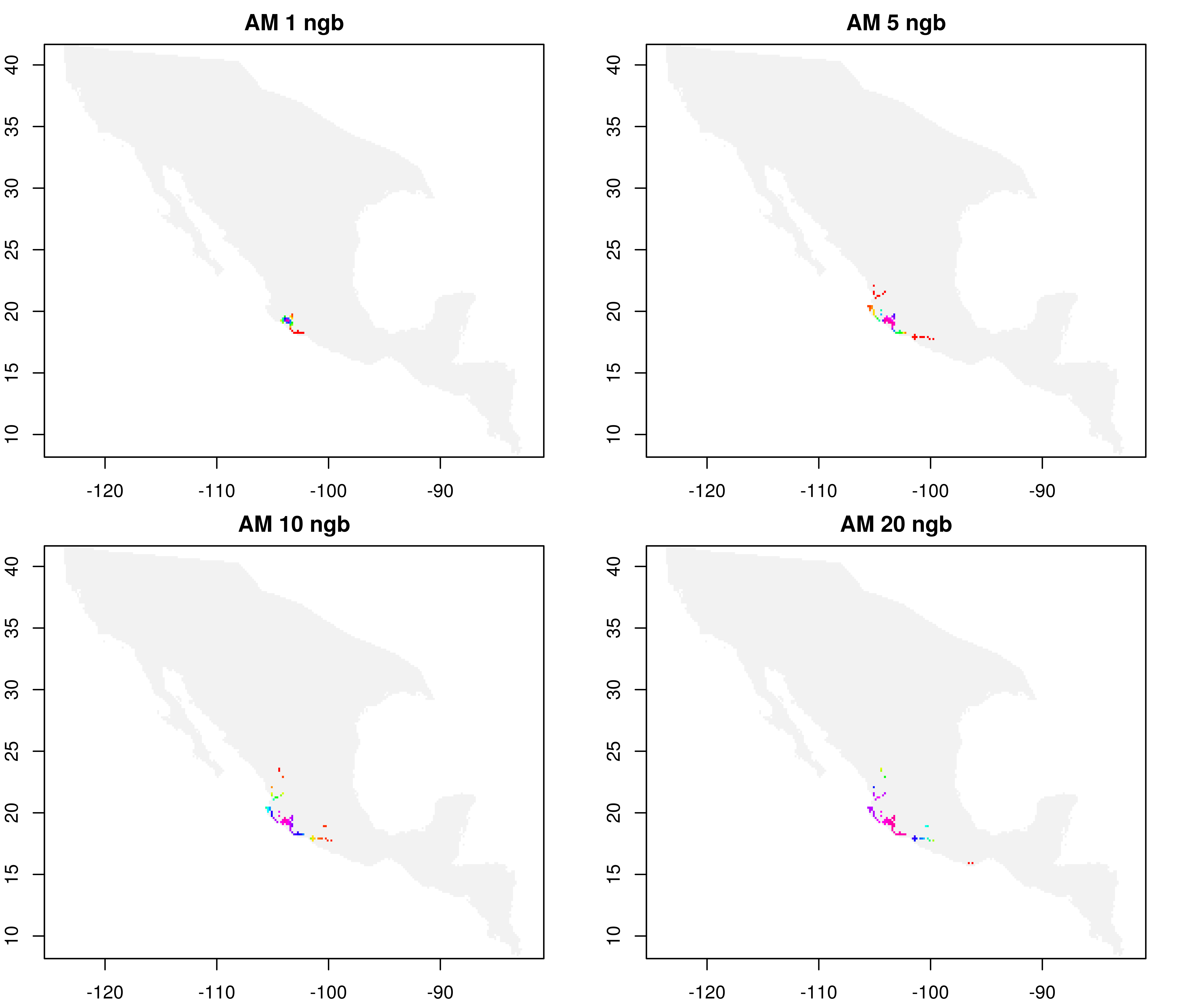}
	\caption{Frequency of visitation to suitable niche cells of \textit{Dismorphia amphione} at 18 km, 36 kms, 90 kms and 360 kms of connectivity.Purple means high frequency of visitation, red low frequency.}
	 \label{fig:AM}
\end{figure}

\begin{figure}[h!]

	\centering
	\includegraphics[scale=.6]{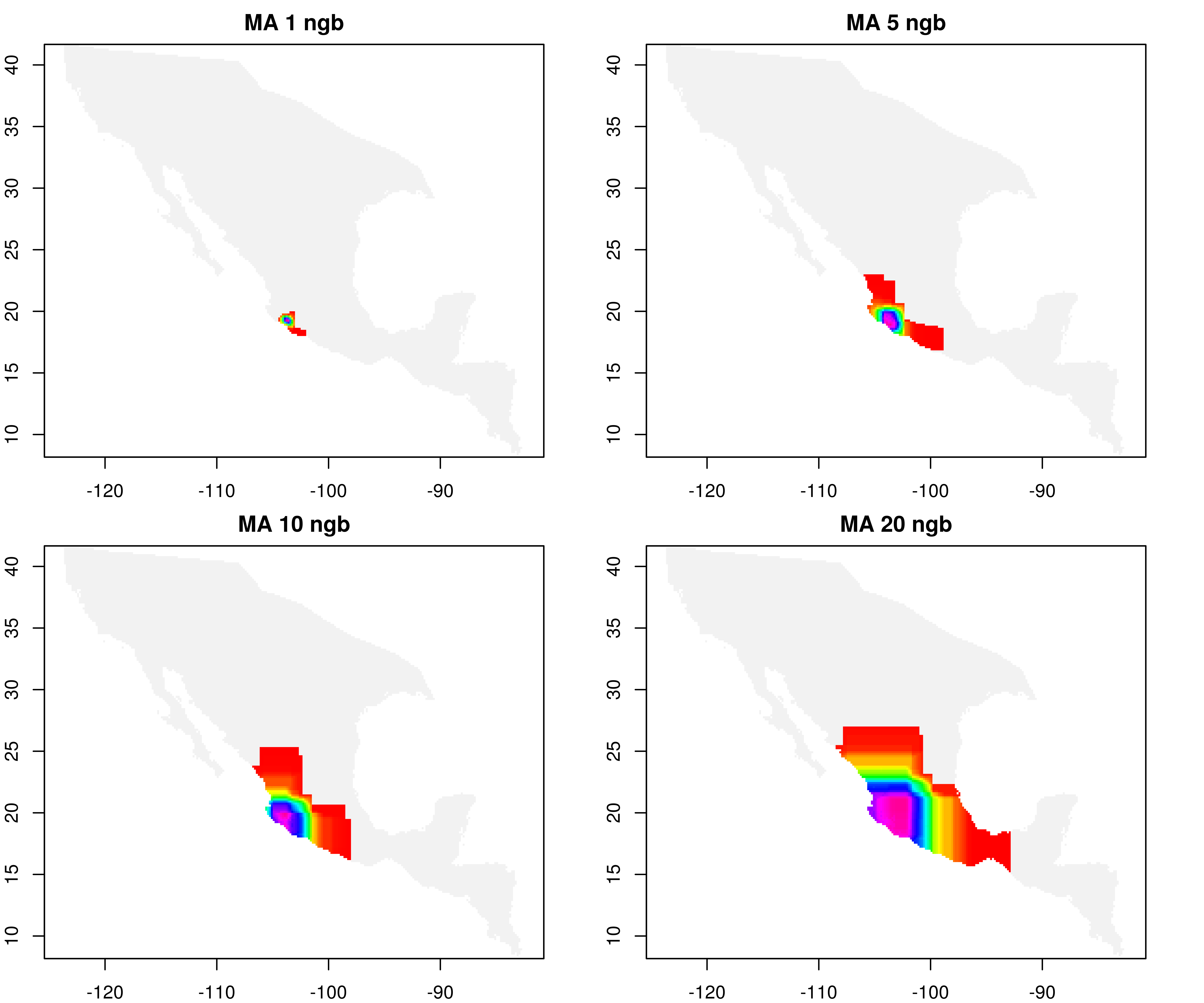}
	\caption{Frequency of visitation from suitable niche cells of \textit{Dismorphia amphione} at 18 km, 36 kms, 90 kms and 360 kms of connectivity.Purple means high frequency of visitation, red low frequency.} 
	\label{fig:MA}
\end{figure}

The first eigenvector of $\mathbf{A}_j(t) \mathbf{C}_j$ therefore indicates how ''visitable" suitable patches are. The first eigenvector of $\mathbf{C}_j\mathbf{A}_j(t)$ indicates how visitable are cells (suitable or not) that can be reached from suitable cells. This may be provide a way to estimate the set of accessible cells, also called $\textbf{M}$ by some authors \citep{Barve2011,Soberon2005}. Figures \ref{fig:AM} and \ref{fig:MA} illustrate the difference.

Finally, as will be discussed later, the matrix $\mathbf{A}_j(t) \mathbf{C}_j$ is singular (in general). This means that it has no inverse, and consequently that there is an ``arrow of time" in equation \ref{eq:automata2}. One can, in general, calculate a final state from an initial one, but from the final state it is not possible, in general, to calculate the initial one.
%

\clearpage



\end{spacing}

\section{Biosketches}

\textbf{Jorge Soberon} is an ecologist interested in the theoretical aspects of biogeographical and macroecological patterns.\\ \textbf{Luis Osorio-Olvera} is an ecologist interested in understanding the biodiversity patterns in time and space. He uses and develops models and computational tools for answering macroecological questions.

\end{document}